\documentclass{elsarticle}
\usepackage{amsmath}
\usepackage{color}
\usepackage[colorlinks=true, citecolor=blue]{hyperref}
\usepackage{bbm}
\usepackage{braket}

\begin{document}

\begin{frontmatter}

\title{Can quantum probes satisfy the weak equivalence principle?}

\author[mainaddress]{Luigi Seveso}

\author[mainaddress,secondaryaddress]{Matteo G. A. Paris}

\address[mainaddress]{Quantum Technology Lab, Dipartimento di Fisica, Universit\`a degli Studi  di Milano, I-20133 Milano,  Italy}
\address[secondaryaddress]{INFN, Sezione di Milano, I-20133 Milano, Italy}

\begin{abstract}
We address the question whether quantum probes in a gravitational field can be considered as test particles obeying the weak equivalence principle (WEP). A formulation of the WEP is proposed which applies also in the quantum regime, while maintaining the physical content of its classical counterpart. Such formulation requires the introduction of a gravitational field not to modify the Fisher information about the mass of a freely-falling probe, extractable through measurements of its position. We discover that, while in a uniform field quantum probes satisfy our formulation of the WEP exactly, gravity gradients can encode nontrivial information about the particle's mass in its wavefunction, leading to violations of the WEP.   
\end{abstract}


\end{frontmatter}

\section{Motivation and methods}
The weak equivalence principle (WEP) is one of the foundational bedrock of classical gravitational theory \cite{weinberg1972gravitation,will1993theory,di2015nonequivalence}. It states that the solutions of the equations of motion for a structureless particle falling in a gravitational field exhibit a special form of universality: they do not depend on the particle's mass. Since the mass represents the charge through which the particle couples to gravity, the WEP suggests that gravity may be fundamentally different from the other forces of nature. In fact, the WEP lies at the basis of the possibility of describing gravity in purely geometric terms \cite{okon11}. 

However, the test bodies that appear in its formulation are just a classical idealization. Physical particles are consistently described only in a quantum framework. The question therefore arises whether a form of universality (i.e. independence from the probe's internal properties) also holds for quantum particles falling under gravity and, if not, how does the principle of equivalence emerges in the classical limit.   

Assessing the validity of the WEP for freely-falling quantum particles offers interesting conceptual challenges \cite{viola1997testing,rabinowitz2006theory,lammerzahl1998minimal,lammerzahl1996equivalence,greenberger1968role,davies04}. In fact, the formulation of the WEP in general relativity refers to test particles with a conserved four-momentum, moving along precise trajectories. However, the description of the dynamics of a quantum particle in terms of a wavefunction is markedly different. First, the wavefunction is not by itself localized, which calls into question the abstraction of a test body in relation to a quantum probe. Second, neither position nor momentum of a propagating wavepacket are well-defined classical variables, but instead represent incompatible observables whose measurements are subject to quantum fluctuations according to the uncertainty principle. Therefore, the Galilean procedure of preparing probes in identical dynamical conditions (same initial position and velocity), and letting them evolve freely, loses operational meaning. The concept of a trajectory dissipates and one can only speak about the results of position measurements. As a consequence, the theory of quantum measurements is expected to play an important part in the formulation of any quantum version of the WEP.

These fundamental difficulties \cite{Sonego95} may be ascribed to the fact that a quantum particle does not follow a unique trajectory, making it challenging to associate a unique geometry to spacetime. In fact, from the viewpoint of the path integral formulation of quantum mechanics \cite{feynman2005quantum}, a particle follows all possible trajectories between two fixed spacetime events. Even if the classical trajectory represents the most important contribution to the total amplitude, fluctuations around it are expected from nearby trajectories. Such fluctuations arise in powers of Planck's constant $\hbar$, so that when $\hbar \to 0$ only the classical trajectory predicted by general relativity survives. The present paper aims to discuss the problem of what becomes of the WEP in the opposite regime, i.e. when quantum fluctuations are turned on. 

Apart for the previously mentioned problems, a further difficulty is linked to the fact that the quantum dynamics of a probe under gravity is often mass-dependent \cite{seveso2016quantum}. One may have thought of identifying universality of free-fall in the quantum regime with mass-independence of the wavefunction. After all, the wavefunction provides a complete description of the physical state of a quantum system and thus plays a role similar to the solution of the equations of motion in the classical setting. However, the mass of a particle appears explicitly in the dynamical evolution equations -- which is in stark contrast with the theory of classical point particles in gravitational fields. 

For example, in the non-relativistic limit, the action for a classical particle of mass $m$ in a Newtonian potential $\varphi$ is 
\begin{equation}
S=\int dt \left(\frac{m \dot{\mathbf{x}}^2}{2}-m\varphi\right)\;.
\end{equation}
The mathematical statement of the universality of free-fall is the fact that $m$ appears only as a multiplicative constant, and thus does not enter the equations of motion. The same happens in a fully relativistic (but still classical) context where the action takes the form 
\begin{equation}\label{relact}
S=-mc\int ds\;,
\end{equation}
the corresponding extremal paths being the geodesics with respect to the metric element $ds^2=g_{\mu\nu}dx^\mu dx^\nu$. However, the action for a field $\phi$ describing a quantum scalar particle of mass $m$ is
\begin{equation}\label{maction}
S = \frac{\hbar^2}{2m}\,\int d_4x \,\sqrt{-g}\,\left(g^{\mu\nu}D_{\mu}\phi^\dagger D_{\nu}\phi+\frac{m^2c^2}{\hbar^2} \phi^\dagger \phi\right)\;.
\end{equation} 
The constant in front has been chosen so that the action has the correct dimensions and agrees with the non-relativistic limit (see also Section \ref{back2}). The corresponding equation of motion 
\begin{equation}\label{kgequation}
\frac{1}{\sqrt{-g}}\partial_\mu(\sqrt{-g}\, g^{\mu\nu}\partial_\nu\phi)+\frac{m^2 c^2}{\hbar^2}\phi=0\;.
\end{equation} 
depends parametrically on the combination $\hbar/m$. Whenever quantum effects become relevant, an explicit dependence of the  dynamics on the particle's mass is unavoidable. Therefore, the wavefunction cannot play the role of the classical trajectory in the formulation of the WEP. 

In the present paper, we approach the problem from an information-theory perspective. The WEP can be interpreted as a statement about the information that an experimenter is allowed to extract about the mass of a freely-falling particle by monitoring its trajectory. That is, the experimenter sets up an experiment with test bodies of different masses, which are left freely-falling under gravity. For test bodies obeying the laws of classical physics, the experimenter can gain no information about the masses of the probes by knowing their trajectories. Can quantumness of the probe allow the experimenter to extract a nonzero amount of information?

To quantify information about an unknown parameter, one has to resort to quantum parameter estimation theory \cite{holevo2003statistical,paris2009quantum,BraunsteinCaves1994}. Since in quantum mechanics the mass of a particle does not correspond directly to an observable, one has to infer its value from the statistics of the measurements of some other observable. We are interested in measurements of position. From the outcomes of such measurements, the experimenter builds an estimator, i.e. a post-processing of the measurements which produces an estimate of the parameter. Being a random variable, any estimator has a mean square error around the true value, which quantifies the sensitivity of any given measurement strategy. One is interested in keeping the average error as small as possible. It is a standard result \cite{cramer1945mathematical} that, for locally unbiased estimators, the mean square error is lower-bounded by the inverse of the Fisher information, multiplied by the number of measurements. More details about the quantum parameter estimation problem can be found in section \ref{back1}. 

Though the Fisher information has a precise statistical meaning, it can be  interpreted more generally as a measure of the information content of a model about an unknown parameter \cite{amari2007methods}. In our case, the statistical model consists of the parametric family of position-space wavefunctions, which are obtained as solutions of the matter wave equation coupled to gravity, labelled by the possible values of the mass $m$.  Measurements of position are then used to extract information on $m$, which is quantified by the Fisher information $F_x(m)$. 

Because quantum dynamics under gravity depends on the ratio $\hbar/m$, the Fisher information $F_x(m)$ is in general nonzero. This however should not be taken as quick proof that the gravitational coupling violates universality in a quantum setting. In fact, $F_x(m)$ would be nonzero even in the absence of any gravitational field. In other words, the dependence on the ratio $\hbar/m$ of matter wave equations like \eqref{kgequation} survives in the limit when the metric becomes flat. For example, a free gaussian wavepacket in non-relativistic quantum mechanics, has a variance $\sigma^2$ which spreads with time like $\sigma^2(t)=\sigma^2(0)+\hbar t/2m$. In principle, one can extract information on $m$ by monitoring the velocity with which the wavepacket spreads. The conclusion is that we must carefully distinguish between the information on the mass which is present when the field is removed, and the information which is explicitly introduced by the gravitational coupling. 

We are now ready to state our formulation of quantum universality of free-fall. For the WEP to apply to quantum probes, we will require the introduction of a gravitational field not to change the Fisher information about their mass when in free-fall, extractable through measurements of their position, that is
\begin{equation}
F_x(m)=F_x(m)|_{free}\;,
\end{equation}  
where $F_x(m)|_{free}$ is the Fisher information in the free case, i.e. when no external field is present. 

The rest of the paper is structured in order to discuss to what degree quantum mechanics is compatible with such a statement. We start, in Section \ref{back2}, by introducing the  relevant background material about quantum estimation theory and the spacetime propagation of a quantum probe. We then focus on two separate testbeds for the WEP. At first, in section \ref{bchnonrelativistic}, we consider the non-relativistic limit (i.e. the first nontrivial term in the $1/c$ expansion) of equation  \eqref{kgequation} and solve its quantum dynamics up to second order in the disentangling of the time evolution operator via the Baker-Campbell-Hausdorff (BCH) formula \cite{serre2009lie}. In this way, we are able to discuss the effects of gravity gradients on the Fisher information and to conclude that while in a uniform field quantum probes satisfy the WEP exactly, gravity gradients can encode nontrivial information about the mass, leading to violations of the WEP. Then, in section \ref{semiclassicalrel} we start again from the relativistic Klein-Gordon equation  \eqref{kgequation} and solve it in the semiclassical approximation (i.e. by considering the first nontrivial term in the $\hbar$ expansion), showing how the WEP is recovered in the classical limit $\hbar\rightarrow 0$.

\section{Background}

\subsection{Parameter estimation theory}\label{back1}
Many tasks in physics take the form of estimating an unknown parameter. A typical problem is to learn a probability distribution from a series of measurements. Such a probability distribution may describe the uncertainty in the measurement of a classical variable due to stochastic noise, or it may describe the intrinsic uncertainty in the measurement of a quantum observable. The experimenter assumes to know its functional form, except for the value of one or more parameters, and has to produce an estimate for them so as to agree as closely as possible with the data. Naturally, her estimate will in general be different from the true value of the parameter. The question is what is the lowest possible error she can achieve, and how can one reach this ultimate limit.   

The problem of parameter estimation is particularly relevant for extracting information from quantum systems. In quantum mechanics, it is often the case that quantities of interest do not directly correspond to an observable, i.e. there is no Hermitian operator whose spectrum gives the possible values of the parameter. A typical example is a Hamiltonian parameter \cite{pang2014quantum}, that is a parameter which appears inside the Hamiltonian, e.g. a coupling constant. 

For definiteness, let us assume the Hamiltonian of the system is $H(\lambda^*)$, where $\lambda^*$ is the true, but unknown, value of the parameter we are interested in. The initial state of the system is given by the density operator $\rho(0)$, which is left evolving unitarily for a time $t$, i.e. $\rho(0)\to e^{-iH(\lambda^*)t}\,\rho(0)\,e^{iH(\lambda^*)t}$. We assume that the initial preparation $\rho(0)$ is independent of the value of $\lambda^*$, i.e. the encoding of the parameter on the system is entirely due to its dynamical evolution. Since $\lambda^*$ is unknown, the experimenter in reality has to discriminate among a whole family of density operators 
\begin{equation}
\rho_\lambda(t) = e^{-iH(\lambda)t}\,\rho(0)\,e^{iH(\lambda)t}
\end{equation} 
where the parameter $\lambda$ belongs to a parameter space $\Lambda$ which includes the true value $\lambda^*$. To this end, she performs a measurement on the system, for example a projective measurement of an observable $\hat X$. Let the outcomes of the measurements be denoted by $x$ (the eigenvalues of $\hat X$). Then the probability of the outcomes is
\begin{equation}\label{statmodel}
p_\lambda(x) = \braket{x|\rho_\lambda(t)|x}\;,
\end{equation}
where $\ket{x}$ are the eigenstates of $\hat X$. The parametric family of probability distributions \eqref{statmodel} defines the statistical model of the problem. 

From the outcomes of the measurement, the experimenter builds an estimator $\hat \lambda$. Being a function of the sampled values $(x_1,\,x_2,\,\dots,\,x_N)\in \chi^N$ ($N$ is the number of repeated measurements of $\hat X$ and $\chi$ the sample space of $\hat X$), the estimator $\hat \lambda$ is itself a random variable. If the estimator is locally unbiased then its expectation value $\langle \hat \lambda\rangle$ equals the true value of the parameter $\lambda^*$, where the expectation is taken with respect to $p_{\lambda^*}$. We are interested moreover in keeping its fluctuations around $\lambda^*$, i.e. $\text{Var}(\hat\lambda)=\langle (\hat\lambda - \lambda^*)^2\rangle$, as small as possible. A lower bound to the value of $\text{Var}(\hat\lambda)$ for any possible locally unbiased estimator is given by the Cram\'er-Rao theorem \cite{cramer1945mathematical},
\begin{equation}
\text{Var}(\hat\lambda)\geq \frac{1}{N\,F_X(\lambda^*)}\;.
\end{equation}
The Fisher information $F_X(\lambda^*)$ in particular is given by the expectation value of the score squared, i.e.
\begin{equation}
F_X(\lambda)= \langle (\partial_\lambda \log p_\lambda)^2 \rangle= \int d\mu(x)\,p_{\lambda}(x)(\partial_\lambda\log p_{\lambda}(x))^2\;,
\end{equation} 
where $\mu$ is the geometric measure on the sample space $\chi$, i.e. the statistical model is normalized such that $\int d\mu(x)\, p_\lambda(x)=1$.

For a pure statistical model, i.e. $\rho(0)=\ket{\psi}\bra{\psi}$, the Fisher information takes the form
\begin{equation}
F_X(\lambda) = \int d\mu({x})\,|\psi_\lambda({x},t)|^2\,(\partial_\lambda \log|\psi_\lambda({x},t)|^2)^2\;, 
\end{equation} 
where $\psi_\lambda(x,t)$ is the wavefunction in the $\hat X$-representation, i.e. $\psi_\lambda(x,t) = \braket{x|e^{-iH(\lambda)t}|\psi}$. Here, we will deal with pure models and with measurements of position, so that the measured observable $\hat X$ will be the position operator $\hat{\mathbf{x}}$ and the unknown parameter will be the mass $m$ of the particle.

\subsection{Propagation of matter waves under gravity}\label{back2}
In this section, we review how to describe the propagation of a quantum probe on a fixed spacetime background. More precisely, we consider the case of a relativistic spinless particle of mass $m$ which moves through a spacetime described by the metric $g_{\mu\nu}$. The particle emerges from the quantization of a relativistic scalar field $\phi$, whose action functional is given by equation \eqref{maction}, where $g = \text{det}(g_{\mu\nu})$, $g^{\mu\nu}$ is the inverse metric tensor and $D_\mu$ is the covariant derivative in the metric background. In \eqref{maction}, minimal-coupling between the matter field and gravity has been assumed, i.e. interaction terms involving the scalar curvature have been neglected.\footnote{In particular, the most relevant such term would be of the form $\xi R\, \phi^\dagger \phi$, where $R$ is the Ricci curvature. The reason for setting $\xi$ to zero is twofold. First, if present, such a term would trivially violate the equivalence principle from the beginning: since the Ricci curvature $R$ involves derivatives of the Christoffel symbols, it would not vanish in a local inertial frame. Moreover, it is in general negligible anyway, as long as the particle propagates through regions much smaller than the curvature length scale.} 

From the action \eqref{maction}, it follows that the field $\phi$ satisfies the covariant Klein-Gordon equation \eqref{kgequation}. In order to give a probabilistic, single particle interpretation to \eqref{kgequation}, one may proceed as in the case of the Schr\"odinger equation, i.e. rewriting it as a continuity equation and identifying the temporal component of the conserved current as a bona fide probability density if nonnegative. Equation \eqref{kgequation} can indeed be rewritten in the form of a covariant continuity equation,
\begin{equation}\label{conteq}
D_\mu (\phi^\dagger\partial^\mu \phi-\phi\, \partial^\mu\phi^\dagger)=0\;.
\end{equation}
The corresponding 4-current, normalized so as to agree with the non-relativistic limit, takes the form
\begin{equation}\label{4current}
j^\mu = \frac{i\hbar}{2m}(\phi^\dagger\partial^\mu \phi-\phi\, \partial^\mu\phi^\dagger)\;.
\end{equation}
From \eqref{conteq}, it follows that, upon integrating over three dimensional space, 
\begin{equation}
\partial_t\,\left(\int d_3x\, \sqrt{-g}\,j^t(x)\right)=0\;.
\end{equation}
Therefore, our candidate to the role of a position-space density is $\sqrt{-g}\,j^t$: if such quantity is normalized to unity at some initial time, it remains so for all times. 

However, $\sqrt{-g}\,j^t$ is not necessarily nonnegative, which in principle forbids any probabilistic interpretation of the relativistic wave equation \eqref{kgequation}. It is indeed well-known \cite{bjorken1964relativistic} that no consistent relativistic single-particle theory can exist:  relativity implies the possibility of pair-production and thus particle number is not conserved. Therefore, we make use of relativistic wave equations like \eqref{kgequation} only as approximations, with the caveat that they be considered at length scales much bigger than the Compton wavelength of the particle.    

Let us now discuss what happens of equation \eqref{kgequation} in the non-relativistic limit. The metric tensor $g_{\mu\nu}$ in the weak-field limit takes the form 
\begin{equation}\label{metric}
g_{\mu\nu} = \eta_{\mu\nu}+\frac{2\varphi}{c^2}\,\delta_{\mu\nu}\;,
\end{equation}  
where $\eta_{\mu\nu}$ is the Minkowski metric, $c$ is the speed of light and $\varphi(\mathbf x)$ is the Newtonian gravitational potential (this follows from solving the Einstein's field equations to order $1/c^2$ \cite{friedlander1975wave}). One can then assume for $\phi$ an ansatz of the form
\begin{equation}\label{ansatz}
\phi(\mathbf{x}, t)= e^{-\frac{imc^2}{\hbar}t}\,\psi(\mathbf{x},t)\;.
\end{equation}  
In the limit $c\to\infty$, the exponential factor in front is rapidly oscillating, while $\psi$ depends on $t$ much more slowly and plays the role of the non-relativistic wavefunction. Substituting the ansatz \eqref{ansatz} inside the Klein-Gordon equation \eqref{kgequation} and collecting similar powers of $1/c$, one obtains \cite{kiefer1991quantum} to lowest nontrivial order the Schr\"odinger-type equation  
\begin{equation}\label{schrodinger}
i\hbar\,\partial_t \psi(\mathbf{x},t)= -\frac{\hbar^2}{2m}\,\psi(\mathbf{x},t) + m\,\varphi(\mathbf{x})\;. 
\end{equation} 
Thus, in the non-relativistic limit the Newtonian potential energy $m\varphi$ enters into the Schr\"odinger equation in the canonical way, despite the fact that gravity is geometrical in origin \cite{sakurai2011modern}.

Finally, let us return to the fully relativistic wave equation \eqref{kgequation} and solve it in the complementary situation when $c$ is kept finite, but $\phi$ is expanded in powers of $\hbar$. That is, we are considering the probe's dynamics to be fully relativistic (with the caveat mentioned before), but amenable to a semiclassical approximation. Explicitly, $\phi$ is taken to obey the ansatz
\begin{equation}\label{semiclansatz}
\phi(x) = A(x)\,e^{\frac{i}{\hbar}S(x)}\;,
\end{equation}
where $A$ and $S$ are real functions of the spacetime coordinates. By substituting in \eqref{kgequation} and separating real and imaginary parts, one obtains the system of equations
\begin{equation}\label{system}
\begin{cases}
\hbar^2 D_\mu \partial^\mu A - A\, \partial_\mu S\, \partial^\mu S +m^2  A=0\;,\\
2\partial_\mu A\,\partial^\mu S + A\, D_\mu \partial^\mu S=0\;.
\end{cases}
\end{equation}

The second equation is strictly equivalent to the continuity equation \eqref{conteq}, since when multiplied by $A$ it takes the form
\begin{equation}\label{semiclconteq}
D_\mu(A^2 \partial^\mu S)=0\;,
\end{equation}
where $A^2\partial^\mu S$ can be checked to be proportional to the 4-current \eqref{4current} using the ansatz \eqref{semiclansatz}. The first equation can be rearranged in the form
\begin{equation}\label{kgsemicl}
\partial_\mu S\, \partial^\mu S -m^2  =\hbar^2 \frac{D_\mu \partial^\mu A}{A}\;.
\end{equation}
An asymptotic solution can be obtained through a WKB expansion truncated to order $O(\hbar)$ \cite{landau1958course}, i.e.
\begin{equation}
A= A_0 + \hbar A_1 +O(\hbar^2)\;,\qquad S = S_0 +\hbar S_1+O(\hbar^2)\;.
\end{equation}
Neglecting terms of order  $\hbar^2$ in  \eqref{kgsemicl} requires that 
\begin{equation}
\left|\hbar^2 \frac{D_\mu \partial^\mu A}{A}\right|\ll m^2\;,
\end{equation}
i.e. that the Compton wavelength of the particle be much smaller than the length scale of variation of $A$. Under this assumption, equation \eqref{kgsemicl} becomes to lowest order
\begin{equation}\label{hjeq}
\partial_\mu S_0\, \partial^\mu S_0 = m^2 \;,
\end{equation}
which is the classical Hamilton-Jacobi equation. Therefore $S_0$ is the classical action and $\partial_\mu S_0$ is the classical 4-mome{}ntum $p_\mu$. To determine $A_0$, one would have to solve the continuity equation \eqref{semiclconteq}, 
\begin{equation}\label{last}
D_\mu(A_0^2\, p^\mu)=0\;.
\end{equation}
Finally, the corrections to the next order are determined by solving the system
\begin{equation}\label{system1}
\begin{cases}
\partial_\mu S_1\, p^\mu=0\;,\\
D_\mu(A_0^2\, \partial^\mu S_1 + 2 A_0 A_1 \, p^\mu)=0\;.
\end{cases}
\end{equation}

\section{The fully quantum, non-relativistic case: effects of gravity gradients on the Fisher information}\label{bchnonrelativistic}
In this section, we focus on the case of a properly quantum probe which moves non-relativistically in a weak external field. Accordingly, we set $\hbar$ to unity, and look for approximate solutions of the Schr\"odinger equation \eqref{schrodinger}. The time evolution operator 
\begin{equation}
U_t= e^{-iHt}= e^{\frac{it\Delta}{2m}-imt\varphi(\mathbf{x})}\;
\end{equation} 
can be disentangled via the BCH formula \cite{serre2009lie}. To second order, one has in general that
\begin{equation}\label{bch}
e^{A+B}\sim e^{\frac{1}{6}(2[B,[A,B]+[A,[A,B]])}\;	e^{\frac{1}{2}[A,B]}\,e^B\, e^A\;.
\end{equation}
In our case, $A=it \Delta/2m$ and $B=-imt\varphi$. One thus finds 
\begin{equation}\label{comm}
\begin{split}
[A,B]=& \frac{t^2}{2}[\Delta, \varphi(\mathbf{x})]= \frac{t^2}{2}(\Delta \varphi(\mathbf{x})+2\nabla\varphi(\mathbf{x})\cdot\nabla)\\=&t^2\,\nabla\varphi(\mathbf{x})\cdot\nabla =- t^2 \mathbf{g}(\mathbf{x}) \cdot\nabla\;.
\end{split}
\end{equation}
To simplify the commutator, one has to use the classical field equation for $\varphi(\mathbf{x})$, i.e. $\Delta\varphi(\mathbf{x})=0$ ($\mathbf{g}$ is the gravitational field $\mathbf{g}(\mathbf{x})=-\nabla\varphi(\mathbf{x})$). Moreover,
\begin{equation}\label{comm1}
[B,[A,B]]=imt^3\,[\varphi(\mathbf{x}),\mathbf{g}(\mathbf{x})\cdot\nabla]=imt^3\,\mathbf{g(\mathbf{x})}^2 \;,
\end{equation}
and
\begin{equation}\label{comm2}
[A,[A,B]]=-\frac{it^3}{2m}\,[\Delta, \mathbf{g}(\mathbf{x})\cdot\nabla]=-\frac{it^3}{m}\,\nabla \mathbf{g(\mathbf{x})}\cdot \nabla\nabla\;,
\end{equation}
which depends explicitly on the gravity gradient $\nabla\mathbf{g}$.\footnote{The notation $\nabla \mathbf{g}\cdot \nabla\nabla$ stands for the second-order differential operator $(\partial_i g_j)\,\partial_i\partial_j$, where repeated latin indices are summed over from 1 to 3. Notice also that $\partial_ig_j=\partial_jg_i$. Thus, in expressions like $\nabla \mathbf{g}\cdot \mathbf{v}$, with $\mathbf{v}$ an arbitrary vector, it is immaterial whether $\mathbf{v}$ is contracted with $\nabla$ or with $\mathbf{g}$.} 

Let us emphasize that formula \eqref{bch} is exact for uniform fields. In fact, all the terms neglected in the BCH expansion involve further commutators with $A$ or $B$ of the previously obtained commutators \eqref{comm1} and \eqref{comm2}. All such commutators vanish identically for uniform fields. Thus, the exact propagator for uniform fields is 
\begin{equation}\label{propuniform}
U_t= e^{\frac{img^2t^3}{3}}\,e^{-\frac{gt^2}{2}\cdot \nabla}\,e^{-imt\varphi(\mathbf{x})}\,U_{t,\,free}\;,
\end{equation}
where $U_{t,free}=\text{exp}(-it\Delta/2m)$ is the time evolution operator in the free case. Equation \eqref{propuniform} can also be rewritten via Hadamard's lemma\footnote{That is, $e^A e^B e^{-A}= e^{\sum_{n=0}^\infty \text{ad}_A^n(B)/n!}$, where $\text{ad}_A(B)=[A,B]$. For a uniform field, only the first two terms of the series  are nonzero.} as 
\begin{equation}
\begin{split}
U_t&=e^{\frac{img^2t^3}{3}}\,e^{-\frac{gt^2}{2}\cdot \nabla}\,e^{-imt\varphi(\mathbf{x})}\,e^{\frac{gt^2}{2}\cdot \nabla}\,e^{-\frac{gt^2}{2}\cdot \nabla}U_{t,\,free}\\&=e^{\frac{-img^2t^3}{6}}\,e^{-imt\varphi(\mathbf{x})}\,e^{-\frac{gt^2}{2}\cdot \nabla}\,U_{t,\,free}
\end{split}
\end{equation}
We have thus recovered a well-known result \cite{kennard1927quantenmechanik,kennard1929quantum}: the propagator in a uniform gravitational field amounts to the composition of a pure phase factor, a translation operator along the classical trajectory, and the free evolution operator. 

The implication is that the information an observer can extract about the mass of a freely-falling probe is the same as in the free case. In fact, let $\psi(\mathbf{x},0)$ describe a wavepacket at time $t=0$ which is left evolving in a uniform field, and let $\psi_{free}(\mathbf{x},t)$ describe its free evolution. The wavefunction for general time $t$ is given by
\begin{equation}
\psi(\mathbf{x}, t) = U_t\, \psi(\mathbf{x}, 0) = e^{\frac{-img^2t^3}{6}}\,e^{-imt\varphi(\mathbf{x})}\,\psi_{free}\left(\mathbf{x}-\frac{\mathbf{g}t^2}{2}, t\right)
\end{equation}
It is now sufficient to use the deviation with respect to the classical trajectory as a new variable of integration in order to see that
\begin{equation}
\begin{split}
F_x(m)=& \int d\mathbf{x}\,|\psi(\mathbf{x},t)|^2\,(\partial_m \log|\psi(\mathbf{x},t)|^2)^2\\ =& \int d\mathbf{x}\,|\psi_{free}(\mathbf{x}-\mathbf{g}t^2/2,t)|^2\,(\partial_m \log|\psi_{free}(\mathbf{x}-\mathbf{g}t^2/2,t)|^2)^2\\=&\int d\mathbf{x}\,|\psi_{free}(\mathbf{x},t)|^2\,(\partial_m \log|\psi_{free}(\mathbf{x},t)|^2)^2\\=& \,F_x(m)|_{free}\,, 
\end{split}
\end{equation}
which is precisely what we mean by universality of free-fall in the quantum regime.

Let us return to the case of an arbitrary gravitational potential $\varphi(\mathbf{x})$. We will solve the Schr\"odinger equation \eqref{schrodinger} operatorially to second order in the BCH expansion \eqref{bch}. We will also make the approximation of neglecting terms at the exponent which are second order in the gravity gradient or its derivatives, but will keep terms which involve first derivatives of the field. In fact, the neglected terms are of the same order as terms already neglected by truncating the BCH expansion and which involve further commutators of expressions \eqref{comm1} and \eqref{comm2}. 
The time evolution operator using \eqref{bch}, \eqref{comm}, \eqref{comm1} and \eqref{comm2} reads
\begin{equation}\label{propagator}
U_t \sim e^{\frac{imt^3}{3}\mathbf{g(\mathbf{x})}^2}\, e^{-\frac{it^3}{6m}\nabla \mathbf{g(\mathbf{x})}\cdot\nabla\nabla-\frac{\mathbf{g(\mathbf{x})}t^2}{2}\cdot\nabla}\,e^{-imt\varphi(\mathbf{x})}\,U_{t,\,free}\;.
\end{equation} 
We have disentangled and rearranged the exponentials, ignoring contributions which would entail higher powers of the gravity gradient or its derivatives.

In analogy with the uniform case, we can manipulate \eqref{propagator} by collecting in front all the phase factors. In \ref{appendix} it is proved that, once this is done, the position-space density reads
\begin{equation}\label{appeq}
|\psi(\mathbf{x},\,t)|^2=\left|\psi_{free}\left(\mathbf{y}+\frac{t^3}{3m}\hat{\mathbf{p}}\cdot\nabla \mathbf{g}(\mathbf{x}), \,t\right)\right|^2\;,
\end{equation}
where 
\begin{equation}
\mathbf{y}(\mathbf{x})=\mathbf{x}+\frac{t^2}{2}(\mathbf{x}\cdot\nabla\mathbf{g}(\mathbf{x})-\mathbf{g}(\mathbf{x}))+\frac{5t^4}{48}\nabla\mathbf{g}^2(\mathbf{x})\;.
\end{equation}
Due to the presence of the second term in parenthesis involving the momentum operator $\hat{\mathbf{p}}$, the functional dependence of the position-space probability density on the mass is different compared to the free case. An experimenter can use repeated measurements of position to extract a higher information on the probe's mass thanks to its gravitational coupling.

If such a term is negligible, however, one may shift variable of integration $\mathbf{x}\to \mathbf{y}(\mathbf{x})$ and then the Fisher information would be unchanged, with a nontrivial sample-space measure $\mu$, i.e. 
\begin{equation}
F_x(m)\sim \int d\mu(\mathbf{x})\,|\psi_{free}(\mathbf{x},t)|^2\,(\partial_m\log |\psi_{free}(\mathbf{x},t)|^2)^2\;.
\end{equation}
We can estimate very roughly when this is possible by substituting the momentum operator with its classical expression in a uniform field, $\mathbf{p}=m\mathbf{g}t$ and imposing the condition
\begin{equation}
\left|\frac{t^3}{m}\mathbf{p}\cdot\nabla\mathbf{g}\right|\ll |\mathbf{g}t^2|\,,
\end{equation} 
which implies $|\mathbf{g}t^2|\ll \ell$, where $\ell$ is the characteristic scale over which the gravitational field is changing. Therefore, as long as the particle does not have the time to explore regions of the order of the curvature scale, universality is recovered. 

In general however the quantum wavefunction is sensitive to gravity gradients, which can imprint nontrivial information about the mass. The quantum wavefunction behaves in analogy with a classical extended body, which in fact can also sense tidal forces and thus does not move in general geodesically \cite{geroch1975motion}. 

\section{The semiclassical relativistic case: recovering the WEP}\label{semiclassicalrel}
In the previous section, we considered a probe which is fully quantum, but non-relativistic, 
and showed that universality is recovered in the limit when the wavefunction is sufficiently 
localized with respect to the curvature scale. In this section, we consider the complementary 
situation where the probe is relativistic, but its dynamics can be treated semiclassically. 
Accordingly, we set $c$ to 1, but keep powers of $\hbar$ explicit. 

We are going to discuss the solutions of the system of equation \eqref{system}, focusing in particular on their functional dependence on the mass $m$. This in turn determines the information about the mass available to any observer, and thus whether the WEP can be satisfied or not by a quantum probe. 

The first equation in \eqref{system}, equation \eqref{kgsemicl}, leads to lowest order in powers of $\hbar$ to a Hamilton-Jacobi equation, see \eqref{hjeq}. Its solution $S_0$ is therefore the classical action \eqref{relact} and $\partial_\mu S_0$ is the classical 4-momentum $p_\mu$. $S_0$ thus depends on $m$ only multiplicatively.

The second equation in \eqref{system} leads to \eqref{last}. Since $p^\mu$ is the classical 4-momentum, it also depends on $m$  multiplicatively, i.e. $p^\mu = m\, dx^\mu /ds$ and $x^\mu$ does not depend on $m$ because of the WEP for classical probes. This means that the differential equation for $A_0$ is actually mass-independent, and so is $A_0$. 

In conclusion, the solution of \eqref{kgequation} to lowest-order in powers of $\hbar$ is
\begin{equation}
\phi(x) \sim A_0(x) \, e^{\frac{i}{\hbar} S_0(x)}\qquad (0^{\text{th}}\;\text{order})\;,
\end{equation}
where $S_0$ is the classical action for a classical point-particle of mass $m$ and $A_0$ is mass-independent. The corresponding probability density is 
\begin{equation}
\sqrt{-g} j^t = -\sqrt{-g}\,\frac{A_0 p^t}{m}\,,
\end{equation}
which is mass-independent and thus leads to vanishing Fisher information. We conclude that the WEP emerges in the geometrical optics limit of quantum mechanics, with the rays being the geodesics of the manifold and the wave surfaces the submanifolds of constant classical action.

Now let us study the corrections arising at the next order in the semiclassical expansion. The system of equations for the corrections of order $\hbar$ is given in \eqref{system1}. One can see that $S_1$ is mass-independent, but the equation for $A_1$ is mass-dependent. It can be made independent of $m$ 
by shifting to the adimensional variable $x^\mu\to mx^\mu$. Since the introduction of the gravitational field modifies the functional form of $A_1$, it also changes nontrivially its functional dependence on the mass, and thus also the Fisher information. Therefore, we conclude again that the universality of free-fall is in general violated when quantum fluctuations around the classical trajectory are turned on. 

\section{Conclusions}
In this paper, we have addressed the question whether quantum probes in a gravitational field can be approximated as test particles obeying the WEP. To this end, we have assumed an information theory-inspired definition of universality of free-fall in the quantum regime. In particular, we have identified the Fisher information about the mass of a freely-falling probe, extractable from position measurements, as the truly universal quantity for quantum free-fall, since it is unchanged by the introduction of a uniform gravitational field. 

Though quantum mechanics respects our formulation of the WEP to lowest order in the BCH expansion of the propagator, this does not hold at higher orders. The quantum wavefunction can sense gravity gradients, which 
encode nontrivial information about the mass both in its phase and in its amplitude, thus leading to an increase of the corresponding Fisher information compared to the free case. We therefore conclude that the WEP is untenable for a quantum particle described by a wavefunction. Our results instead agree with a general conclusion common to many quantum gravity programs, i.e. the fact that in the quantum regime no unique geometry may be associated to the spacetime structure of our universe \cite{Sonego95,okon11}.

Finally, we discussed how the WEP is recovered in the classical limit. The WEP is recovered either when the particle is localized with respect to the curvature scale, and inhomogeneities in the gravity field can thus be neglected, or in the semiclassical limit $\hbar\to 0$, when quantum fluctuations around the geodesic trajectory are suppressed. 

\appendix
\section{Proof of equation \eqref{appeq}}
\label{appendix}

In this appendix we are going to provide more details about the derivation of equation \eqref{appeq}. Starting from equation \eqref{propagator} for the propagator, we employ Hadamard's lemma to rewrite
\begin{equation}
e^{-\frac{it^3}{6m}\nabla \mathbf{g(\mathbf{x})}\cdot\nabla\nabla-\frac{\mathbf{g(\mathbf{x})}t^2}{2}\cdot\nabla}\,e^{-imt\varphi(\mathbf{x})}\,e^{\frac{it^3}{6m}\nabla \mathbf{g(\mathbf{x})}\cdot\nabla\nabla+\frac{\mathbf{g(\mathbf{x})}t^2}{2}\cdot\nabla}
\end{equation}
as
\begin{equation}
\sim e^{-imt\varphi{(\mathbf{x})}+\frac{t^4}{6}\nabla\mathbf{g}^2(\mathbf{x})\cdot \nabla-\frac{im\mathbf{g}^2(\mathbf{x})t^3}{2}}\;,
\end{equation}
which can be further disentangled as
\begin{equation}
\sim e^{-\frac{im\mathbf{g}^2(\mathbf{x})t^3}{2}}\,e^{-\frac{imt}{2}\nabla\mathbf{g}^2(\mathbf{x})\cdot\mathbf{g}(\mathbf{x})}\,e^{-imt\varphi(\mathbf{x})}\, e^{\frac{t^4}{6}\nabla\mathbf{g}^2(\mathbf{x})\cdot  \nabla}\;.
\end{equation}
Thus, the propagator can be rewritten with all derivative operators acting on the right as
\begin{equation}
e^{-\frac{im\mathbf{g}^2(\mathbf{x})t^3}{6}}\,e^{-\frac{imt}{2}\nabla\mathbf{g}^2(\mathbf{x})\cdot\mathbf{g}(\mathbf{x})}\,e^{-imt\varphi(\mathbf{x})}\, e^{-\frac{it^3}{6m}\nabla \mathbf{g(\mathbf{x})}\cdot\nabla\nabla+\frac{t^4}{6}\nabla\mathbf{g}^2(\mathbf{x})\cdot  \nabla-\frac{\mathbf{g(\mathbf{x})}t^2}{2}\cdot\nabla}\,U_{t, \,free}\;.
\end{equation}
Let us denote by $\mathcal{D}$ the derivative operator which appears in the last exponential,
\begin{equation}\label{derop}
\mathcal{D}=-\frac{it^3}{6m}\nabla \mathbf{g(\mathbf{x})}\cdot\nabla\nabla+\frac{t^4}{6}\nabla\mathbf{g}^2(\mathbf{x})\cdot  \nabla-\frac{\mathbf{g(\mathbf{x})}t^2}{2}\cdot\nabla
\end{equation}
Contrary to what we found for uniform fields, in the presence of gravity gradients $\mathcal D$ is no longer a simple displacement. This is because of the first term in \eqref{derop} which is second order in derivatives. 

When $U_{t,\,free}$ acts on the initial wavepacket $\psi(x, 0)$, one obtains 
\begin{equation}
U_{t,\,free}\, \psi(x, 0)= \psi_{free}(x, t) = e^{x\cdot\nabla}\psi_{free}(0,t)= e^{-\mathcal D}(e^{\mathcal D}e^{x\cdot\nabla}e^{-\mathcal D})\,\psi_{free}(0,t)\;.
\end{equation} 
The term in parenthesis can be transformed again using Hadamard's lemma. The time-evolved wavefunction can finally be written as
\begin{equation}
\psi(\mathbf{x}, t) \sim e^{-\frac{im\mathbf{g}^2(\mathbf{x})t^3}{6}}\,e^{-\frac{imt}{2}\nabla\mathbf{g}^2(\mathbf{x})\cdot\mathbf{g}(\mathbf{x})}\,e^{-imt\varphi(\mathbf{x})}\,\psi_{free}\left(\mathbf{y}(\mathbf{x})+\frac{t^3}{3m}\hat{\mathbf{p}}\cdot\nabla \mathbf{g}(\mathbf{x}), \,t\right)
\end{equation}
where we defined
\begin{equation}
\mathbf{y}(\mathbf{x})=\mathbf{x}+\frac{t^2}{2}(\mathbf{x}\cdot\nabla\mathbf{g}(\mathbf{x})-\mathbf{g}(\mathbf{x}))+\frac{5t^4}{48}\nabla\mathbf{g}^2(\mathbf{x}) 
\end{equation}
and $\hat{\mathbf{p}}=-i\nabla$ is the momentum operator. In particular, the position-space density is
\begin{equation}
|\psi(\mathbf{x},\,t)|^2=\left|\psi_{free}\left(\mathbf{y}+\frac{t^3}{3m}\hat{\mathbf{p}}\cdot\nabla \mathbf{g}(\mathbf{x}), \,t\right)\right|^2\;,
\end{equation}
which is equation \eqref{appeq}.

\bibliographystyle{elsarticle-num}
\bibliography{newref}

\end{document}